\documentclass[conference]{IEEEtran}
\usepackage{amsmath,amsthm}
\usepackage{url}
\usepackage{graphicx}
\usepackage{epstopdf}
\usepackage{tikz}
\usepackage{moreverb}
\usepackage{pdfpages}
\usepackage{multicol}
\newtheorem*{defn}{Definition}

\begin{document}
	
\title{On the linkability of Zcash transactions}
\author{
	\IEEEauthorblockN{Jeffrey Quesnelle}
	\IEEEauthorblockA{University of Michigan-Dearborn}
}
\maketitle

\begin{abstract}
	Zcash is a fork of Bitcoin with optional anonymity features. 
	While transparent transactions are fully linkable, shielded transactions use zero-knowledge proofs to obscure the parties and amounts of the transactions. 
	First, we observe various metrics regarding the usage of shielded addresses.
	Moreover, we show that most coins sent to shielded addresses are later sent back to transparent addresses.
	We then search for round-trip transactions, where the same, or nearly the same number of coins are sent from a transparent address, to a shielded address, and back again to a transparent address.
	We argue that such behavior exhibits high linkability, especially when they occur nearby temporally.
	Using this heuristic our analysis matched 31.5\% of all coins sent to shielded addresses.
\end{abstract}

\section{Introduction}
Zcash is a cryptocurrency which provides a mechanism to obscure the source, destination, and amounts of transactions. 
These transactions are known as \emph{shielded} transactions.
The privacy of shielded transactions is achieved by using zk-SNARKs, which are a type of zero-knowledge proof \cite{BSCG+14}.

The use of shielded transactions is optional.
As with Bitcoin, Zcash coins are sent to addresses, which can be thought of as public keys.
Only the owner of the corresponding private key can generate a new transaction to transfer the coins.
In Zcash there are two classes of addresses: transparent and shielded. 
When a transaction occurs between transparent addresses the transfer is akin to Bitcoin transactions, where the parties involved and the amounts are fully visible.
Only when a transaction occurs between two shielded addresses are Zcash's privacy guarantees in full effect.

Generating shielded transactions comes at a significant computational cost, taking between thirty seconds and one minute to create on a modern desktop computer. 
Perhaps because of this, most Zcash transactions do not involve shielded addresses. 
In Section \ref{sec:shielded} we quantify this, finding that only about 3.5\% of all Zcash coins are controlled by shielded addresses on average.

Many transactions involving shielded addresses exhibit a particular pattern: first, coins are sent from a transparent address to a shielded address. 
Soon after, an identical or nearly identical amount of coins are sent back to a transparent address. 
We call such transactions \emph{round-trip transactions} and argue that the controlling party is likely the same for both transactions. 
In Section \ref{sec:roundtrip}, we show that 31.5\%\footnote{All statistics in this paper were calculated on the Zcash blockchain ending in block 196304, which was mined on October 4, 2017} of all coins sent to shielded addresses are likely involved in a round-trip transaction, potentially removing their unlinkability.
Lastly, we discuss these results in Section \ref{sec:conclusion}.

\section{Background}

Zcash retains nearly all of the original Bitcoin functionality.
Transactions between transparent addresses (also known as \emph{t-addrs}) are essentially equivalent in form to Bitcoin transactions.
In Bitcoin, coins are transferred by referencing the outputs of previous transactions and providing a digital signature that proves ownership of the address the coins were sent to.
Since coins are only created as part of the block reward for miners, it is straightforward to audit the correctness of every transaction: simply trace back each transaction output, verifying digital signatures along way back to a block reward. 

However, the radical simplicity of Bitcoin's transparent ledger has significant privacy drawbacks.
If an address is ever associated with a real-world identity it becomes trivial to trace the source and destination of all the user's coins. 
This loss of privacy can be mitigated by never re-using the same address, although even the original Bitcoin whitepaper notes that linking will remain unavoidable \cite{Nak09}.
Several studies of the anonymitity of Bitcoin have been undertaken, showing that analysis can overcome even a dedicated effort at obfuscation \cite{RH11,MPJ+13}.
Thus, when transferring between two t-addrs in Zcash, the transaction is fully linkable.

Unlike Bitcoin, Zcash uses a second type of address known as shielded addresses (also known as \emph{z-addrs}). 
Transactions involving z-addrs are carried out in the JoinSplit structure, which is a new structure added to the Bitcoin transaction format. 
A JoinSplit contains three essential fields: the number of coins entering the shielded pool (known as \texttt{vpub\_old}), the number of coins exiting the shielded pool (known as \texttt{vpub\_new}), and a field holding a zero-knowledge proof attesting to the legitimacy of the transaction.

There are three different operations that a JoinSplit may perform (see Figure \ref{fig:joinsplittypes}). 
The first is a \emph{shielding transaction}, where coins are sent from a t-addr to a z-addr.
This corresponds with a non-zero \texttt{vpub\_old}.
Thus, in shielding transactions the amount being sent to a z-addr is public, but the z-addr itself is not.
The second operation is a \emph{de-shielding transaction}, where coins are sent from a z-addr to a t-addr.
This corresponds with a non-zero \texttt{vpub\_new}.
Likewise, the z-addr remains private but the amount is public.
The final operation is a \emph{shielded transaction}, where coins are transferred between two z-addrs. 
For shielded transactions both addresses are private, as well as the amounts.
 
\begin{figure}
	\caption{The three types of operations performed by a JoinSplit}
	\centering
	\begin{tabular}{| l | l | l | l |}
		\cline{2-4}
		\multicolumn{1}{c|}{} & \textbf{Shielding} & \textbf{De-shielding} & \textbf{Shielded} \\ \hline
		Source & t-addr   & z-addr   & z-addr  \\ \hline
		Destination & z-addr   & t-addr   & z-addr  \\ \hline
		Amount & Public   & Public   & Private  \\ \hline
	\end{tabular}
	\label{fig:joinsplittypes}
\end{figure} 
 
\section{Shielded Transactions} \label{sec:shielded}

The anonymity of z-addrs is the fundamental differentiator between Zcash and Bitcoin.
We examine several metrics of this anonymity as it pertains to real-world usage of Zcash.

\subsection{Transaction metrics}

\begin{figure}[h!]
	\centering
	\caption{The distribution of the percentage of transactions in each block that contain at least one JoinSplit}
	{\input{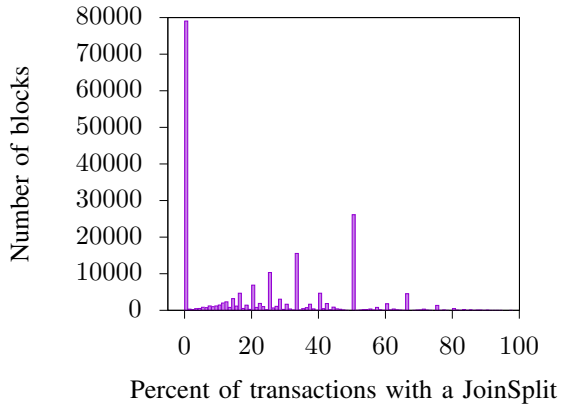}} 
	\label{fig:percentjoinsplits}
\end{figure} 

Overall, 19.6\% (259,220) of all Zcash transactions contained a JoinSplit operation, with 40.2\% of blocks containing no transactions with JoinSplits (see Figure \ref{fig:percentjoinsplits}).
Thus, 80.4\% of all Zcash transactions are trivially linkable back at least one transaction.

Of the 19.6\% of transactions that do contain a JoinSplit, we further classified the transactions into the types of JoinSplit operations being performed.
Every JoinSplit may contain an unknown number of shielded transactions, so there is no way to place an upper bound on the number of shielded transactions.
However, only 1.9\% (5,450) of all JoinSplits have no shielding or deshielding portion, while 40.6\% (116,743) of JoinSplits included a shielding operation and 57.5\% (165,394) included a deshielding operation\footnote{No JoinSplit included both a shielding and a deshielding operation. It is unclear if the Zcash software allows this}.
This evidence supports our eventual hypothesis that most coins sent to z-addrs are sent back to t-addrs. 

\subsection{Coin metrics}

\begin{figure}[h!]
	\centering
	\caption{Size of shielded pool compared to total supply of Zcash coins}
	{\input{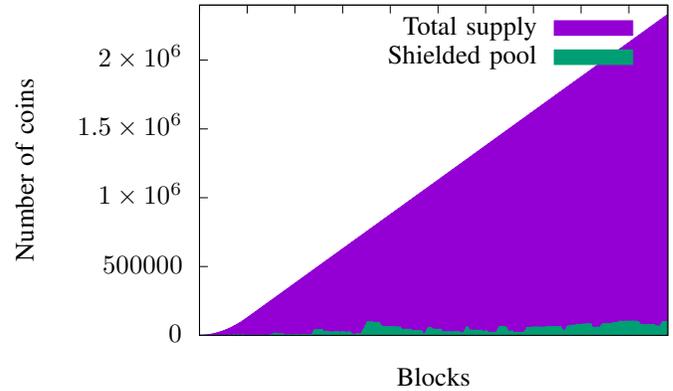}} 
	\label{fig:shieldedtotalsupply}
\end{figure} 

We also investigated the participation of coins in shielded transactions. 
As with Bitcoin and other cryptocurrencies, transactions are grouped into blocks, and miners must solve a difficult proof-of-work problem on the block and announce it to the network.
Valid blocks reward the miner with a certain number of coins in what is known as the block reward.
Thus, with each block the available supply grows slightly, until some fixed\footnote{21 million for Bitcoin and Zcash} cap.

In Zcash, coins exist either in the \emph{transparent pool}, where they are controlled by a t-addr, or the \emph{shielded pool}, controlled by a z-addr.
To determine the size of the shielded pool, the difference between \texttt{vpub\_old} and \texttt{vpub\_new} of each JoinSplit is calculated.
This is the net differential into the shielded pool. 
The total pool size is kept as a running sum of the differential for each block.
Our results (see Figure \ref{fig:shieldedtotalsupply}) show that the shielded pool only ever contains a relatively small percentage of the overall supply of Zcash coins\footnote{For coin totals, only the whole number portion of coins are reported}.
For the final block cataloged, the percentage of coins in the shielded pool was 4.3\%.
The average percentage per block was 3.5\%.

\subsection{Discussion}

Taken together, the previous results suggest an environment where transactions using z-addrs are infrequently used.
In an environment where shielded transactions (i.e. z-addr to z-addr and fully private) are commonplace, it is reasonable to assume that many to most JoinSplits would contain no shielding or deshielding operation.
However, out of the 287,587 JoinSplits cataloged, only 5,450 met these conditions. 
Furthermore, only a relatively small percentage of coins are ever present in the shielded pool.
Together, these two conditions lead us to hypothesize that when coins are introduced to the shielded pool, they are soon returned back to the transparent pool.

Why are z-addr transactions so rare? 
We suggest that the reason is likely practical: most wallet programs do not support z-addrs.
According to the Zcash community website\footnote{\url{https://www.zcashcommunity.com/wallets/} accessed on November 1, 2017} no web-based wallets support z-addrs.
In addition, all cryptocurrency exchanges that trade Zcash only accept t-addrs.
For users that wish to take advantage of the privacy features afforded by z-addrs while still maintaining the practical utility of t-addrs, a simple solution would be to send coins to a z-addr, and then send them back to a t-addr.
This would have the effect of removing the linkability of the transactions.
However, because the amounts of the shielding and deshielding operations are public, these types of transactions may still be linkable.

\section{Round-trip Transactions} \label{sec:roundtrip}

In this section we use the public amount of shielding and deshielding operations to build a candidate set of linkable transactions that pass through the shielded pool.
The official Zcash website notes\footnote{\url{https://z.cash/blog/transaction-linkability.html} accessed on November 2, 2017} that such analysis is possible.
However, we believe this study is the first to perform the analysis and show the prevalence of such transactions. 

\subsection{Defining}

\begin{defn}
	Let $j_1$ and $j_2$ be JoinSplits included in blocks of height $h_1$ and $h_2$ with $h_2 > h_1$. If there exists exactly one pair of transactions $(t_1,t_2)$ with $j_1 \in t_1$, $j_2 \in t_2$, and $\texttt{vpub\_new}(j_2) = \texttt{vpub\_old}(j_1)$, then we say $(t_1, t_2)$ is a \emph{round-trip transaction} (RTT).
\end{defn}

RTTs are an ordered pair of transactions containing JoinSplits where the shielding amount in the first transaction is equal to the deshielding amount in the second transaction. 
In addition, the second transaction must appear in a later block than the first transaction. 
Lastly, we are only concerned about those pairs where there is exactly one pair that fits the criteria, e.g. if $t_1$ shields $10.1234$ coins, we are looking if there is exactly one JoinSplit in any later block that deshields $10.1234$ coins.

We note that although there is strong circumstantial evidence that $t_1$ and $t_2$ in a RTT are linked, the conclusion is not definitive. 
Not all RTTs identified by our heuristic are actually linked; the match may be coincidental.

\subsection{Methodology}

To detect RTTs, we first modified an existing Bitcoin blockchain parser to support Zcash. 
We then used this to build a relational database of the blockchain, linking blocks, transactions, and JoinSplits.
Figure \ref{fig:query} gives the pseudocode for a query to find JoinSplits that meet the criteria for RTTs.

\begin{figure}[h!]
	\caption{Pseudocode for SQL query to find potential round-trip transactions}
\begin{verbatimtab}[2]
SELECT *
FROM JoinSplits E,F
(
	SELECT *
	FROM JoinSplits G,H
	WHERE G.vpub_old > 0 
		AND G.vpub_old = H.vpub_new
		AND H.Block > G.Block
	GROUP BY G.vpub_old
	HAVING COUNT(G.Id) = 1
) I
WHERE E.vpub_old = F.vpub_new 
	AND E.vpub_old = I.vpub_old
\end{verbatimtab}
	\label{fig:query}
\end{figure}

\subsection{Results}

Our analysis found 10,075 transaction pairs that met our criteria (see Figure \ref{fig:roundtripresults}), transferring a total of 919,220 coins. 
The total number of all coins entering the shielded pool was 2,911,734.
Thus, we determined that 31.5\% of all coins entering the shielded pool were involved in a RTT.

We could not conclusively determine that all RTTs are linked transactions.
However, there is strong circumstantial evidence that the false positive rate of our heuristic is low, and that most of the transactions are indeed linked.
96\% (9673) of our matches involved transactions which appeared within two hours of each other on the blockchain.
In addition, by the definition of RTTs, the amounts being shielded and unshielded are globally unique amongst the entire history of Zcash.
Given the high divisibility\footnote{A Zcash coin can be divided out to 8 decimal places} of Zcash coins, we believe that a single exact match of the shielding and deshielding amounts occurring within a few hours is strong evidence that the transactions are linked.

\begin{figure}[h!]
	\caption{Round-trip transactions by block time difference in minutes}
	\centering
	\begin{tabular}{ r | r | r }
		$\Delta$ block time & \# RTT  & $\Sigma$ coins\\ \hline
		$[0, 5)$ & 1373 & 156,237\\ \hline
		$[5, 15)$ & 5022 & 421,021 \\ \hline
		$[15, 30)$ & 1479 & 147,546\\ \hline
		$[30, 60)$ & 1015 & 95,034\\ \hline
		$[60, 120)$ & 500 & 35,741\\ \hline
		$[120, 1440)$ & 284 & 60,518 \\ \hline
		$[1440, \infty)$ & 402 & 3,120
	\end{tabular}
	\label{fig:roundtripresults}
\end{figure}

\begin{figure}[h!]
	\caption{Top JoinSplits by \texttt{vpub\_old} that are part of a round-trip transaction}
	\centering
	\begin{tabular}{ r | r | r }
		Top $n$ JoinSplits & \# RTT & $\Sigma$ coins \\ \hline
		10 & 10 & 34,153 \\ \hline
		50 & 49 & 143,924\\ \hline
		250 & 236 & 500,163\\ \hline
		500 & 460 & 765,212 \\ \hline
		1000 & 585 & 834,301	
	\end{tabular}
	\label{fig:topjoinsplits}
\end{figure}

Large denomination transfers were particularly likely to be RTTs (see Figure \ref{fig:topjoinsplits}).
Of top 250 JoinSplits by shielding amount, 236 were part of a RTT.
Upon further investigation\footnote{The t-addrs of Zcash miners are known} it was discovered that many of these large JoinSplits were Zcash mining pools sending their block rewards to a z-addr before distributing the coins to the t-addrs of miners.
In this case, miners may be under the impression that source of their coins is private since they are receiving their payout from the pool via a z-addr.
However, because the pool is engaging in RTTs, we were able to determine the t-addrs of the pool's members.

\subsection{Fee-adjusted RTTs}

\begin{figure}
	\caption{Most common fees for non-coinbase transactions}
	\centering
	\begin{tabular}{ l | r | r }
		\multicolumn{1}{c|}{Fee} & \multicolumn{1}{c|}{\# tx} & \multicolumn{1}{c}{\%} \\ \hline
		0.0001     & 523,036 & 46.40\% \\ \hline
		0.001      &  34,203 &  3.03\% \\ \hline
		0.0002     &  33,662 &  2.99\% \\ \hline
		0.00009    &  30,400 &  2.70\% \\ \hline
		0.00005    &  24,127 &  2.14\% \\ \hline
		0.00000226 &  23,679 &  2.10\% \\ \hline
		0		   &  16,154 &  1.43\% \\
	\end{tabular}
	\label{fig:commonfees}
\end{figure}

When creating a Zcash transaction, the sender may choose to offer a small amount of Zcash as an incentive for miners to include the transaction in a block. 
This is known as a \emph{fee}.
Although fees are not required, 98.6\% of all non-coinbase\footnote{The first transaction in a block is the \emph{coinbase} transaction, which specifies the receiver of the block reward. Coinbase transactions do not have fees} Zcash transactions include a fee.
Figure \ref{fig:commonfees} shows the most common fees.
The fee used most frequently was 0.0001 Zcash, which 46.4\% of transactions used.

For RTTs, $\texttt{vpub\_new} = \texttt{vpub\_old}$. 
However, if the party performs any shielded transactions (z-addr to z-addr) before transferring back to the transparent pool, the \texttt{vpub} fields may not match, since the shielded transactions may have also paid fees.
We relax our definition of an RTT to $\texttt{vpub\_new} = \texttt{vpub\_old} - f$ where $f$ is some combination of common fees. We call such transactions \emph{fee-adjusted round trip transactions}.

\begin{figure}
	\caption{1-fee round-trip transactions}
	\centering
	\begin{tabular}{ l | r | r }
		\multicolumn{1}{c|}{Fee} & \multicolumn{1}{c|}{\# RTT} & \multicolumn{1}{c}{$\Sigma$ coins} \\ \hline
		0.0001     &  85 & 1,278 \\ \hline
		0.001      & 149 & 1,360 \\ \hline
		0.0002     & 143 & 1,400 \\ \hline
		0.00009    &   2 &    20 \\ \hline
		0.00005    &   9 &    20 \\ 
	\end{tabular}
	\label{fig:onefeertt}
\end{figure}

For fee-adjusted RTTs we considered only transactions appearing within 24 hours of each other.
In addition, to limit false positives only the five most common fees were used.
We first searched for 1-fee RTTs, which attempts to detect the following pattern: t-addr $\rightarrow$ z-addr $\overset{\text{fee}}{\rightarrow}$  z-addr $\rightarrow$ t-addr.

Figure \ref{fig:onefeertt} gives the results for $1$-fee RTTs.
A total of 388 such transaction pairs were found, accounting for a total of 6,058 coins.
Given the apparent scarcity\footnote{Recall that only 1.9\% (5,450) JoinSplits have no shielding or deshielding operation} of shielded transactions, it is unsurprising that relatively few were found.

\begin{figure}
	\caption{2-fee round-trip transactions}
	\centering
	\begin{tabular}{ l | r | r }
		\multicolumn{1}{c|}{Fee} & \multicolumn{1}{c|}{\# RTT} & \multicolumn{1}{c}{$\Sigma$ coins} \\ \hline
		0.002      & 146 & 1,305 \\ \hline
		0.0012     & 131 & 1,264 \\ \hline
		0.0011     &  29 &   167 \\ \hline
		0.00109    &   0 &     0 \\ \hline
		0.00105    &   2 &    18 \\ \hline
		0.0004     & 137 & 1,316 \\ \hline 
		0.0003     &  19 &   111 \\ \hline 
		0.00029    &   3 &    30 \\ \hline 
		0.00025    &   4 &    30 \\ \hline 
		0.00019    &   7 &    80 \\ \hline 
		0.00018    &   1 &    10 \\ \hline
		0.00015    &   3 &    40 \\ \hline
		0.00014    &   3 &    40 \\
	\end{tabular}
	\label{fig:twofeertt}
\end{figure}

We also searched for 2-fee RTTs, attempting to detect the following:  t-addr $\rightarrow$ z-addr $\overset{\text{fee}}{\rightarrow}$ z-addr $\overset{\text{fee}}{\rightarrow}$ z-addr $\rightarrow$ t-addr.
We considered combinations of the five most common fees, except for those which summed to a common fee (e.g. $0.0001 + 0.0001 = 0.0002$). A total of 485 transaction pairs were found, accounting for a total of 4,411 coins.

\section{Discussion and Conclusion} \label{sec:conclusion}

Bitcoin's transparent ledger allows the source and destination of coins to be traced by any third party. 
To prevent this linking, Zcash employs shielded transactions to obscure these details.
However, only transfers between two z-addrs are truly private.
Empirical evidence suggests that most usage of z-addrs involves shielding or deshielding operations, where the amount transferred is still public.
We have shown that a third party can use this information to link entries and exits from the shielded pool.
In our experiments we were able to identify 31.5\% of all shielded pool coins as likely being involved in round-trip transactions.

When privacy features are optional, users often take the path of least resistance. 
Given the large computational costs of shielded transactions, they are relatively rare.
Round-trip transactions may be an effort to ``have the best of both worlds", but if used incorrectly they do not deter a determined attacker from linking transactions.
To be entirely sure that the source of coins cannot be traced, a user must perform a second, fully shielded transaction after receiving coins in a shielding operation.
If they wish to return the coins to the transparent pool, they must also take care to leave some portion remaining in the shielded pool to avoid performing a RTT.

Lastly, we highlight a real-world case where RTTs can lead to a serious privacy loss.
Many of the coins involved in RTTs come from mining pools distributing the block reward proportionally to users based on their contribution to the pool.
Several of the popular Zcash pools perform an RTT before actually distributing the reward.
It may be that the pool's members do not wish for it to be known that they are engaged in mining.
By virtue of receiving coins from a z-addr, they may believe that the source is obscured.
However, by identifying RTTs, the true source of their coins is revealed, which may have serious repercussions.

\appendix

\subsection{Sample round-trip transactions}
\parindent0pt
$t_1$: \texttt{a2c9f7ad3b1993c40e692da61966f8633d85cb96c07b8810c6b14493978f2b46} \\
$t_2$: \texttt{ab3b717b85a64541c6d4bb2da8c0806da9666fa1979e0f640c7f49c44fea3bca} \\
\texttt{vpub\_old}: 3479.51898254 \\
\texttt{vpub\_new}: 3479.51898254 \\
$\Delta$ block time: 2 minutes \\
\\
$t_1$: \texttt{d4e0047df31d0e1c8a7d311064314a74c43d0677ffcc430f8d093bb1867dd21b} \\
$t_2$: \texttt{b63f4948b405b91c28bd59affc06e12aa8e126cb1f101ab36e1114ee882bb0b3} \\
\texttt{vpub\_old}: 12.14981195 \\
\texttt{vpub\_new}: 12.14981195 \\
$\Delta$ block time: 3 minutes \\
\\
$t_1$: \texttt{a6c87c8e2f20b729a33fec7031b2ead3ec6a001e4aa4c575207c44f2690870e4} \\
$t_2$: \texttt{9f300ecfdfb6a8658f34bd469d74f401dd7233d7a610cb91faaeb4a2b3fdc299} \\
\texttt{vpub\_old}: 3.77326919 \\
\texttt{vpub\_new}: 3.77326919 \\
$\Delta$ block time: 928 minutes \\
\\
$t_1$: \texttt{709e38ab58148f6b2a3eb56621ea502790270386b7c6648baf06a510cf48efaa} \\
$t_2$: \texttt{9f300ecfdfb6a8658f34bd469d74f401dd7233d7a610cb91faaeb4a2b3fdc299} \\
\texttt{vpub\_old}: 220.01805591 \\
\texttt{vpub\_new}: 220.01805591 \\
$\Delta$ block time: 15 minutes \\

\subsection{Sample 1-fee round-trip transaction}
\parindent0pt
$t_1$: \texttt{2641aeece9df50c5275b692a20da6f900a1a42440adc454765d7f3e6a1b1aeef} \\
$t_2$: \texttt{4d83b22ab6967c83f11e4cb6f417623c553364ddc5c8d658027356bc28fa6f1a} \\
\texttt{vpub\_old}: 0.67209594 \\
\texttt{vpub\_new}: 0.67199594 \\
$f$: $0.0001$ \\
$\Delta$ block time: 8 minutes \\

\subsection{Sample 2-fee round-trip transaction}
\parindent0pt
$t_1$: \texttt{84a11d9794e0eb318327dd960b7bfa4e1146855fcb1f0aaf6eb40ceadaf9ecbb} \\
$t_2$: \texttt{855e94b007d66f1ee283374c91b559d02fa397079d6f9b5b9012a668680efd71} \\
\texttt{vpub\_old}: 6.3805 \\
\texttt{vpub\_new}: 6.3794 \\
$f$: $0.0001 + 0.001 = 0.0011$ \\
$\Delta$ block time: 35 minutes \\

\end{document}